\documentclass[aip,apl,amsmath,amssymb,reprint]{revtex4-1}
\usepackage[colorlinks,bookmarks=false,citecolor=blue,linkcolor=blue,urlcolor=blue]{hyperref}
\usepackage[all]{hypcap}   
\usepackage{soul}
\usepackage{placeins}
\usepackage{ulem} 
\usepackage{amsmath,amssymb}
\usepackage{graphicx}
\graphicspath{{figures/}{/}}
\usepackage{dcolumn}
\usepackage{bm}
\usepackage[capitalise,compress]{cleveref}
\crefname{section}{Sec.}{Secs.}
\Crefname{section}{Section}{Sections}

\usepackage{verbatim}
\usepackage{color}
\usepackage{soul}
\usepackage{placeins}  
\usepackage{flafter}     
\usepackage{color}
\usepackage{hhline}
\usepackage{physics}
\usepackage{hyperref}
\usepackage{mathptmx}
\usepackage{etoolbox}
\usepackage{float}
\usepackage{caption}
\usepackage{subcaption}
\usepackage{lipsum} 

\usepackage[mathscr]{eucal}
\usepackage{tikz}
\usepackage{xcolor}

\newcommand{\bo}{\begin{outline}}
\newcommand{\eo}{\end{outline}}

\newcommand{\qed}{\nobreak \ifvmode \relax \else
      \ifdim\lastskip<1.5em \hskip-\lastskip
      \hskip1.5em plus0em minus0.5em \fi \nobreak
      \vrule height0.75em width0.5em depth0.25em\fi}
\begin{document} 

\title{Upper bounds on charging power and tangible advantage in quantum batteries}

\author{Sreeram PG}
\email{sreerampg7@gmail.com}
\affiliation{Department of Physics, Indian Institute of Science Education and Research, Pune 411008, India}
\affiliation{MIT Art Design and Technology University,
Pune 412201, India}
\author{J. Bharathi Kannan}
\email{bharathikannan1130@gmail.com}
\affiliation{Department of Physics, Indian Institute of Science Education and Research, Pune 411008, India}
\author{M. S. Santhanam}
\email{santh@iiserpune.ac.in}
\affiliation{Department of Physics, Indian Institute of Science Education and Research, Pune 411008, India}


\begin{abstract}
Quantum battery is expected to outperform its classical counterpart due to quantum effects. Usually, in a quantum battery made of $N$ cells, quantum advantage is demonstrated through super-extensive scaling of the upper bound to the charging power with $N$. In this work, we show that potential quantum advantage as measured by the power bounds need not translate to {\it tangible} advantage in practice. We demonstrate this by considering an all-to-all coupled spin-chain model of a quantum battery with 2-local interactions. It exhibits super-extensive charging when analyzed using the upper bound derived from the uncertainty principle. Unlike the previously studied models, the contribution to this apparent quantum advantage is two-fold -- arising from both the battery and the charger. The model is also experimentally friendly, as it does not require global couplings and yet generates genuine multipartite entanglement. However, we demonstrate that the potential quantum advantage in this scenario is not tangible by employing a tighter upper bound on power. Additionally, we show that even this tighter bound can fail in a range of physical situations and indicate a quantum enhancement that is intangible in practice. Hence, we argue that actual power transferred must be evaluated along with proper characterization of the resources before claiming quantum advantage.

\end{abstract}
\maketitle

Rapid advancements in quantum information and technologies have sparked an interest in using quantum effects to enhance the performance of energy storage devices or quantum batteries \cite{campaioli2024colloquium, quach2023quantum}. A quantum battery is essentially a quantum system that can be manipulated through unitary operations to store and release energy \cite{alicki2013entanglement,le2018spin,campaioli2024colloquium, maillette2023experimental,santos2019stable,dou2022highly,ahmadi2024nonreciprocal,quach2023quantum}, with the expectation that entanglement and superposition might aid in achieving better performance compared to the standard electro-chemical batteries. In this context, quantum advantage arises in the super-extensive charging power of quantum batteries, {\it i.e.,} the charging power increases super-linearly with the number of energy units. In contrast, the charging speed of classical batteries is independent of size. The anticipated advantage has fuelled extensive research towards realising quantum batteries that can display super-extensive charging property. 

With considerable progress in theoretical frameworks \cite{campaioli2024colloquium, quach2023quantum}, experimental quantum battery devices are beginning to emerge. In 2022, super-extensive quantum battery based on Dicke model was demonstrated using an ensemble of two level systems (realised in organic semiconductor) coupled to an optical mode in a microcavity \cite{QuaMcgGan2022}. A cluster of nuclear spins in an NMR platform had demonstrated quantum advantage in the charging process \cite{JosMah2022}. A superconducting quantum battery was realized using transmon qutrit to characterize its charging and the self-discharging processes \cite{HuQiuSou2022, gemme2024qutrit}, while IBM superconducting transmon chips in Armonk processor could also work as quantum battery \cite{GemGroFer2022}. Other superconducting circuits have emerged as well including quantum phase batteries using doped nano-wire with unpaired-spin surface states \cite{StrIorDur2020} and Xmon qutrit setup \cite{Zheng_2022}.  The Dicke quantum batteries \cite{QuaMcgGan2022} suffer from fast discharge rates, and a recent experiment shows how to enhance energy storage lifetime by a factor of $10^3$ using molecular triplet states \cite{TibGasvan2025}. {
A room-temperature Dicke-type quantum battery using organic microcavity \cite{hymas2025experimental} displays superextensive charging and metastable energy storage. Another realization is based on quantum discord in a copper carboxylate complex \cite{cruz2022quantum}.}

Against this backdrop, as more experimental realizations emerge in different physical platforms, it is necessary to obtain platform independent bounds on charging power of quantum batteries. Consider a quantum battery given by Hamiltonian $H_B$, and it is charged by a charger represented by the Hamiltonian $H_C(t)$. A quantum battery consisting of $N$ {\it globally coupled} cells (spins) can charge $N$ times faster due to quantum correlations among the cells. This must be contrasted with the {\it classical} scenario of charging them independently using $N$ chargers \cite{binder2015quantacell}.  Though quantum correlations aids speedup, in practice, collective charging with complex global correlations is necessary for a potentially extensive ($\sim N$) advantage \cite{gyhm2022quantum}.

Though ergotropy or the maximal unitarily extractable work \cite{allahverdyan2004maximal, garcia2020fluctuations, sone2021quantum, tirone2021quantum, alicki2013entanglement, pg2025dichotomy, touil2021ergotropy} in a quantum battery plays an important role, it is well appreciated that quantum advantage emerges in the battery charging process \cite{hovhannisyan2013entanglement} and not in the energy stored or extracted. Hence, this leads us to consider charging power, {\it  i.e.,} the energy deposited per unit time \cite{quach2023quantum} defined as
\begin{equation}
P(t)=\mathrm{Tr}[(\rho'-\rho)H_B]/t,
\end{equation}
where $\rho'$ is the time evolved state until the charging time $t$. The quantum battery is charged by a charger with Hamiltonian $H_C(t)$. The upper bound on the instantaneous charging power can be obtained from the uncertainty relation as \cite{julia2020bounds}, 
\begin{equation}
P(t)^2 \leq 4 ~ {\Delta H_B(t)^2 ~ \Delta H_C(t)^2},
\label{eq:ubound1}
\end{equation}
where $\Delta H^2$ represents the variance of Hamiltonian $H$. Power bound can capture potential quantum advantage because larger entanglement typically leads to enhanced non-local charging process. 

\begin{figure}[t]
    \centering
    \includegraphics[width=0.5\linewidth]{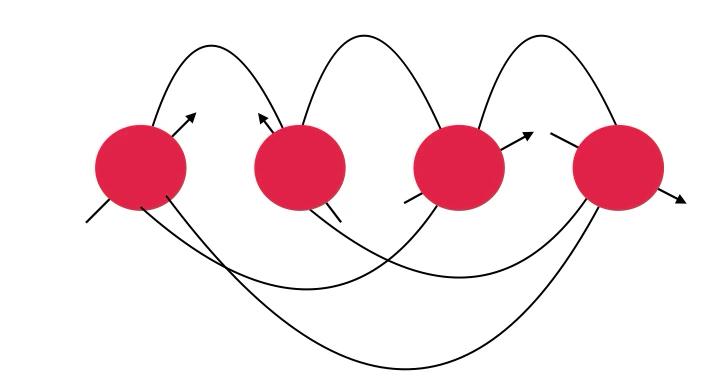}
    \caption{A schematic of an all-to-all interacting spin chain model with 2-local interactions. The chain shows only $N=4$ spins for clarity.}
    \label{fig:spinchain}
\end{figure}

Proper scaling of the energy resource $H_C$ is crucial, as undue advantage can arise even when $H_C$ is extensive. Hence, under a fair physical constraint -- specifically, a limit on the maximum available energy or strength of the charging field -- quantum advantage does not manifest as a superlinear scaling of power with $P \sim N^\alpha$ with $\alpha>1$, but rather as a dramatic reduction in the resources required to achieve the classical scaling of power ($P \sim N$) \cite{julia2020bounds}. A more physically meaningful and tighter bound \cite{julia2020bounds}
\begin{equation}
P(t)^2 \leq  ~ {\Delta H_B(t)^2 ~I_E(t)}. 
\label{fi_bound}
\end{equation}
was obtained by replacing $\Delta H_C(t)^2$ by the Fisher information $I_E(t)$.
Physically, $I_E(t)$ captures the ``speed'' of the battery state evolution in the energy eigenspace by quantifying the probability current, and significantly makes the power bound independent of $H_C$.
While $H_C$ drives the occupancy in energy space, larger $\Delta H_B^2 ~\Delta H_C^2$ in collective charging only indicates the a potential, not necessarily a guaranteed and tangible,  quantum advantage. {If the system shows superlinear charging under realistic conditions even after accounting for the resources needed for coherent charging (as opposed to classical parallel charging), then the superlinear part indicates {\it tangible} advantage.}
While these bounds {discussed above} are physically well-founded and can even be reached in certain cases, they also may not accurately represent the underlying energy dynamics in specific situations. Though the bounds are theoretically elegant and insightful about the mechanism of the quantum advantage, one should perform a more detailed analysis, as the upper bounds as a standalone measure can be quite loose or even misleading for the system at hand.
We find that computing the bound $\Delta H_B^2 ~\Delta H_C^2$ suggests that the system can exhibit a near-optimal quantum charging advantage despite the lack of global interactions. However, upon further scrutiny, we find that the battery state's evolution in the energy basis does not show faster charging with increasing $N.$

To probe a host of such interconnected issues related to the quantum advantage stated by the power bounds (such as in Eqs. \ref{eq:ubound1}, \ref{fi_bound}) as against tangible quantum advantage, in this article, we study an all-to-all coupled spin-chain quantum battery with 2-local interactions (two-body interactions). 
\begin{figure}
    \centering
    \includegraphics[width=0.95\linewidth]{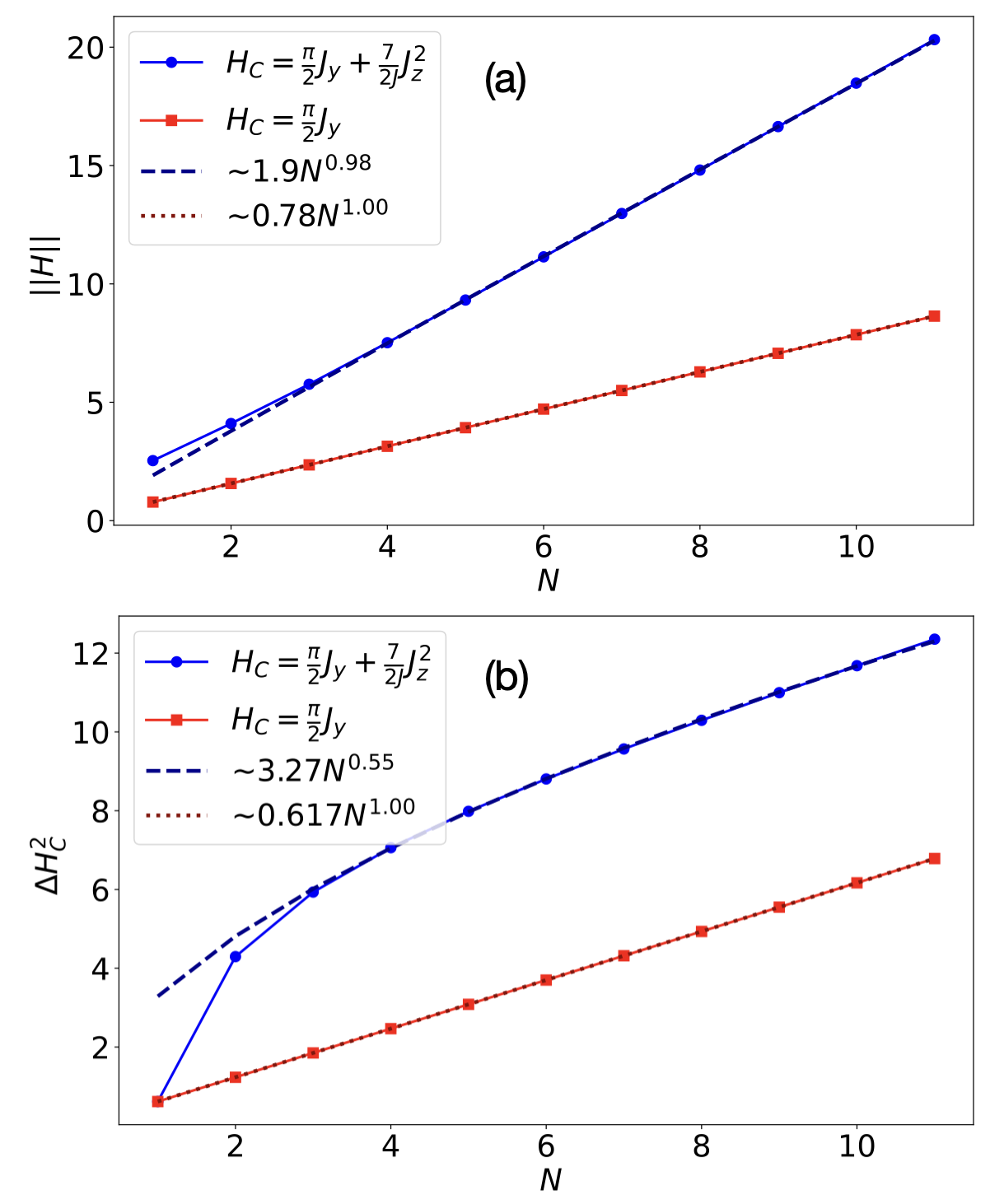}
    \caption{(a) Mean eigenvalue and (b) variance of the eigenvalues of $H_C$ at the kicks ($H_C= \frac{\pi}{2} J_y+ \frac{7}{2j}J_z^2$) and in between the kicks ($H_C= \frac{\pi}{2} J_y$). }
    \label{fig:scaling}
\end{figure}
The spin-chain quantum battery consists of a battery Hamiltonian given by
\begin{equation}
H_B= J_z=\sum_{i=1}^{N} \sigma_z^i,
\label{eq:battery1}   
\end{equation}
and a charger Hamiltonian $H_C$ represented by a collection of spins to which instantaneous kicks of strength $\beta$ are imparted at periodic intervals of $\tau$ \cite{haake1987classical,chaudhury2009quantum, krithika2019nmr}:
\begin{equation}
    H_C= \frac{\pi}{2} J_y + \beta \frac{J_z^2}{2j} \sum_n \delta(t-n \tau).
\label{eq:charger1}
\end{equation}
In this, the angular momentum operators are composed of smaller spins \cite{ghose2008chaos,dogra2019quantum,sreeram2021out}, { i.e.}, $J_{\alpha}= \sum_{i=1}^N \sigma_{\alpha}^i,$ 
where $\alpha=x,y,z$ with $N$ being the number of spins in the chain {such that $j=N/2$}, and $\sigma_{\alpha}$ are the Pauli matrices for spin-$\frac{1}{2}$.
In terms of individual spins, the Hamiltonian $H_C$ can be explicitly written as
\begin{equation}
    H_C= \frac{\pi}{2} \sum_{i=1}^N \sigma_y^i+ \beta \frac{\left(\sum_{i=1}^N \sigma_z^i\right)^2}{2j} \sum_n \delta(t-n \tau). 
\label{charger}
\end{equation}
In $H_C$, the term $J_z=\left(\sum_{i=1}^N \sigma_z^i \right)^2$ leads to coupling of each spin with every other spin via two-body interactions as shown in Fig. \ref{fig:spinchain}.  
During the time interval between consecutive kicks, only $J_y=\sum_{i=1}^N \sigma_y^i$ term is present, leading to a continuous precession. At the instance of a kick, the impulsive nonlinear torsion is turns on, whose strength is determined by $\beta.$ { Physically, such delta kicks are realized differently experimental platforms \cite{chaudhury2009quantum, neill2016ergodic,krithika2019nmr}.} {In NMR experiments \cite{krithika2019nmr} with nuclear spins, kicks are applied using short RF pulse, while nonlinear rotation was realized using spin-spin couplings. In the superconducting qubit system \cite{neill2016ergodic}, the delta kick is implemented through tunable transmon couplings.  In cold atom experiments with Cs atoms \cite{chaudhury2009quantum}, delta kicks were implemented using short magnetic field pulse.} Usually, the kicks could be modelled by a strong square pulse lasting $\epsilon \rightarrow 0$ in time, with the strength $f(t) \sim 1/\epsilon$ acting periodically. 
{ Therefore, at the kicks, the physical Hamiltonian becomes} $ H_C= { \frac{\pi}{2} J_y + \beta \frac{J_z^2}{2j}  \times f(t).}$  Since $f(t)$ does not vary with system size, we may as well set $f(t)=1$ to check the extensivity of the Hamiltonian. 

Figure  \ref{fig:scaling} displays the norm and variance of $H_C$. At instants when kicks are applied, $H_C= \frac{\pi}{2} J_y + \beta \frac{J_z^2}{2j}$ with $\beta=7$, and during the time interval between consecutive kicks $H_C= \frac{\pi}{2} J_y$. Note that $\beta=7$ is the fully chaotic regime for $H_C$. It is clear from Fig. \ref{fig:scaling}(a) that the kicked spin chain Hamiltonian $H_C$ is extensive at the instants of kicks and in between kicks. As the numerical results show in Fig. \ref{fig:scaling}(a), $J_y$ and $J_z^2/2j$ scale linearly with $N$, and therefore the mean energy growing linearly ($\sim N$) is to be expected.  

The variance of $H_C$ is shown in Fig. \ref{fig:scaling}(b). The eigenvalue variance (between successive kicks) of $\frac{\pi}{2}J_y$ scales linearly because of its composite structure being a sum of Pauli matrices and the resulting degeneracies.  For any given quantum number $m$ (corresponding to projection along $z$-axis), there are $\binom{N}{j+m}$ possible configurations where $m \in \{-j, -j+1, \dots, j-1, j\}$. This is unlike the case of single particle ($N=1$), in which $J_y$ operator will have no degeneracies, and its variance would scale quadratically.
Now assume that $\beta \gg 0$ where the kicking term dominates, so that the Hamiltonian effectively reduces to $H_C \approx  \frac{J_z^2}{2j}$ at the instants of kicks. The energy eigenvalue of $\frac{J_z^2}{2j}$ is $E=\frac{m^2}{2j}$.  The $J_z$ operator, with a composite structure as a sum of $N$ Pauli operators, has a similar degeneracy structure as that of $J_y$. Under this circumstance, the variance $\Delta J_z^2$ only scales quadratically $\sim N^2$, whereas in the corresponding single particle case  $\Delta J_z^2$ scales as $\sim N^4$. Furthermore, the $1/j^2$ factor makes $ \left(\Delta \frac{J_z^2}{2j} \right)^2= \frac{1}{4j^2} (\left\langle J_z^4\rangle -\langle J_z^2\rangle^2 \right) $ independent of $N$. These foregoing arguments explain the sublinear scaling of the $\Delta H^2_C$ at the instants of kicks for $\beta=7$, where both $J_y$ and $J_z^2$ terms contribute (Fig. \ref{fig:scaling}(b)).

\begin{figure}
     \centering
     \includegraphics[width=0.45\textwidth]{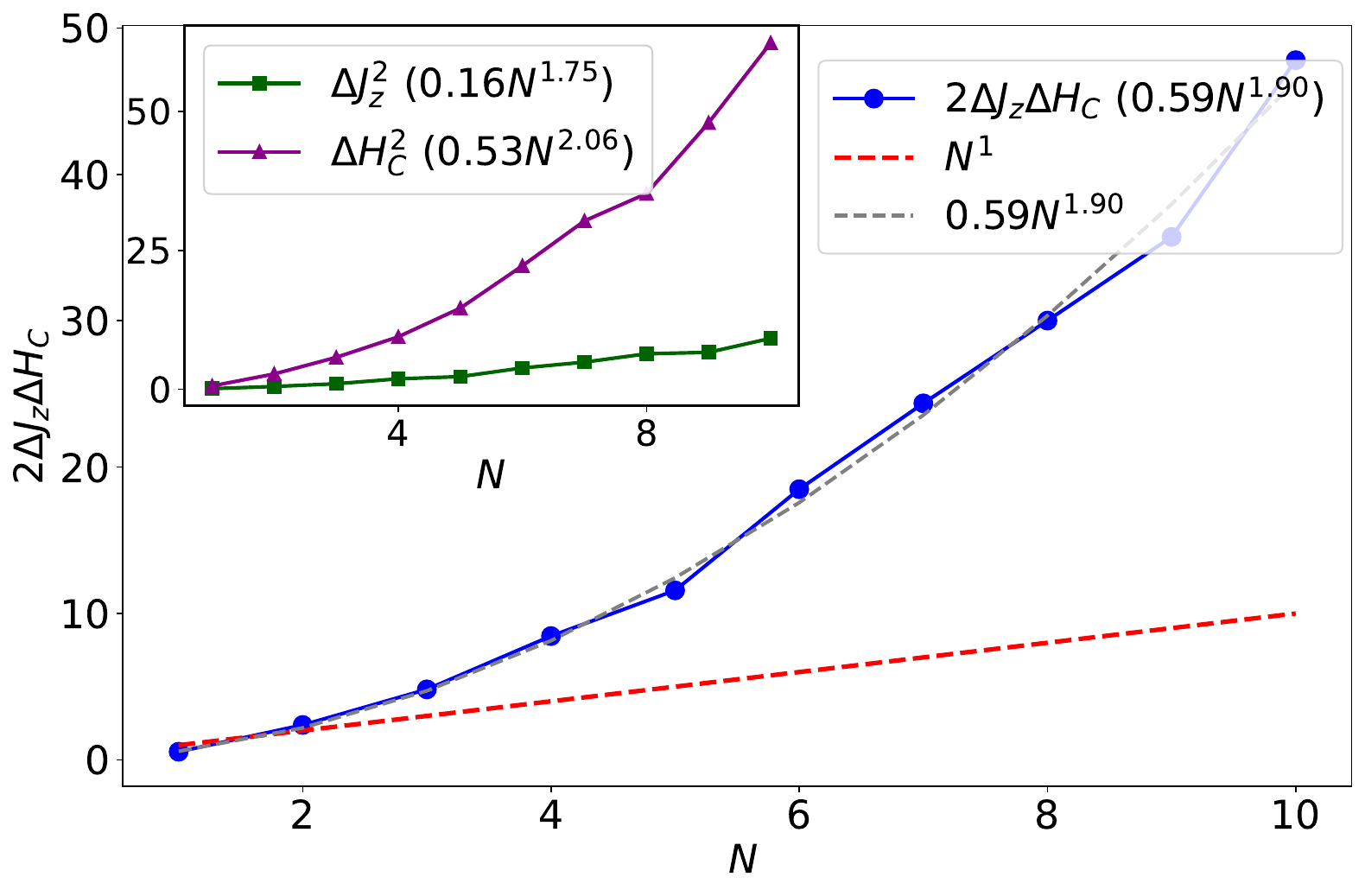}
     \caption{Power scaling in the spin chain battery with the periodically kicked charging, averaged over 50 time steps, is close to quadratic ($\sim N^{1.9}$). The $N^1$ line shows the classical scaling.  {The inset shows the individual scalings of the battery and the charger variances.}}
     \label{fig: powerscaling}
\end{figure}
 
 Having characterised the charger, for further analysis, the internal battery $H_B$ is chosen as in Eq. \ref{eq:battery1}. Note that $[H_C, H_B] \ne 0$, and $H_B$ has an additive structure, { i.e.}, $H_B$ is composed of $N$ independent cells. Since the charger is a periodically kicked system of Eq. \ref{eq:charger1}, time evolution reduces to stroboscopic tracking of the dynamics.  The evolution of the battery state over one time step is obtained from Eq. \ref{charger} as 
 \begin{equation}
     U_C(\tau)= \exp\left(-\frac{\pi}{2}\sum_{i=1}^N \sigma_y^i\right) \exp\left(\frac{\beta}{2j} \left(\sum_{i=1}^N \sigma_z^i\right)^2\right).
 \end{equation}
 The battery charging time is fixed to be 50 time steps,  { i.e.}, 50 applications of $U_C(\tau)$. We choose the initial state to be the fully discharged state of $H_B$, given by the coherent state \cite{haake1987classical}:
 \begin{equation}
     \ket{\theta,\phi}= \exp\{i\theta (J_x \sin\phi- J_y \cos\phi)\} ~ \ket{j,j},
 \end{equation}
 with $(\theta= \pi, \phi=0)$, and $\ket{j,j}$ denotes a simultaneous eigenvector of $J^2$ and $J_z$, with all the spins pointing up. 

Let us first estimate the power bound given by Eq. \ref{eq:ubound1}. For this, we choose the kick strength of  $\beta=7$, well in the fully chaotic regime, and set the initial state to be the coherent state at $(\theta=\pi, \: \phi=0)$, the fully discharged state with respect to $H_B$. As the battery is charged,   the inset of Fig. \ref{fig: powerscaling} shows that the time-averaged variance  of the battery Hamiltonian scales as $\Delta H_B^2 = \Delta J_z^2 \sim N^{1.78}$, and $\Delta H_C^2 \sim N^{2.06}$. In this, the exponent values were estimated through a numerical regression. Combining these results, the power bound scales as $P=2 \Delta {J_z} ~\Delta {H}_C =0.59 ~N^{1.9}$ (Fig. \ref{fig: powerscaling}).  It reveals almost close to quadratic scaling.
Surprisingly, this near-quadratic scaling has been achieved without using any global interactions in the charger.
The maximum bound on power is quadratic in $N$, and is achievable only with a globally interacting charger of the form $H_C^\#= \lambda \bigotimes_{i=1}^{N} h_C^i,$ where $h_C^i$ is a local operator on the $i^\textrm{th}$ spin. Physically, global operation of the form $H_C^\#$ leads to genuine multipartite entanglement among the cells, and can give rise to quadratic scaling.

Despite lacking globally interacting charger, how did we achieve quadratic scaling? This can be explained as follows.
The charger variance $\Delta H_C^2$, computed with respect to the evolved states, is superextensive (inset of Fig. \ref{fig:scaling}). An initial state evolves into highly nonclassical states involving significant quantum correlations, and the variance grows superlinearly, even though the bare Hamiltonian satisfies the thermodynamic consistency condition.  In the charger, as $\beta$ increases, $\Delta H^2_C$ grows increasingly superlinear,  indicating the quantum origin of such superextensivity.  This, combined with the superextensive variance of $H_B$ is the origin of the quantum advantage. This suggests that we can potentially achieve a quadratic scaling without global interactions! Moreover, the 2-local coupling discussed in this model is also experimentally friendly compared to the fragile global interactions.  Together, this sounds sensational. This requires further analysis. Now, a series of different scenarios are considered below to understand the import of the power bounds.

\begin{figure}
    \centering
    \includegraphics[width=0.4\textwidth]{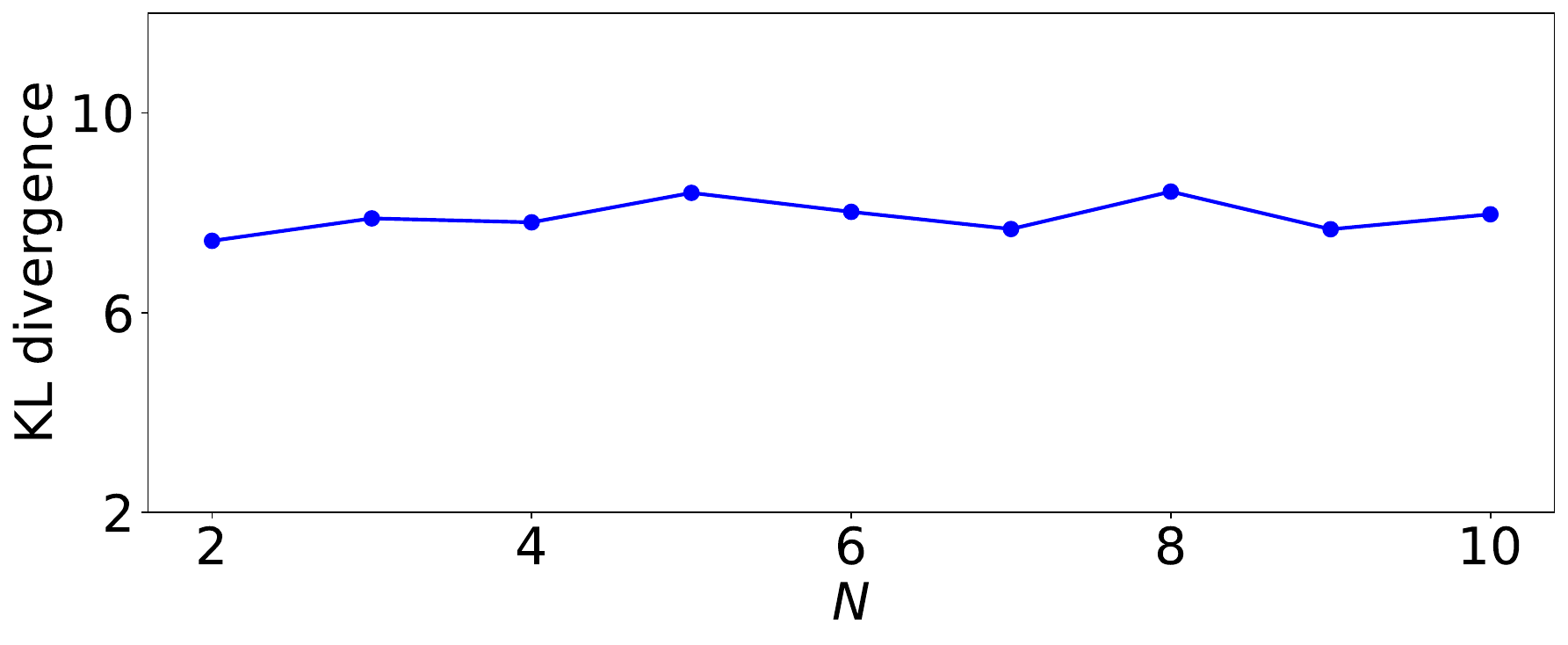}
    \caption{KL divergence averaged over 50 time steps, as a function of the number of spins.}
    \label{fig:FI scaling}
\end{figure}

Now, we recompute the power bound using Fisher Information $I_E$ as given by Eq. \ref{fi_bound}\cite{julia2020bounds}. The battery Hamiltonian in diagonal form is $H_B= \sum E_k ~P_k$, where $E_k, k=1,2, \dots N$ are the energy eigenvalues, and $P_k$ the corresponding eigenvector projectors. For the battery state $\rho(t)$ with projection $p_k(t)= \mathrm{Tr}(P_k ~\rho(t))$,  we have
\begin{align}
     I_E(t)&=\sum_k \left(\frac{d}{dt}  \mathrm{log}_2(p_k(t)) \right)^2 ~ p_k(t), \label{fi} \\
     &\approx \frac{2 ~D_\mathrm{KL}(\bar p(t+\delta t)||\bar{p}(t))}{\delta t^2}.
 \end{align}
 Here $D_\mathrm{KL}(.||.)$ is the Kullback-Leibler (KL) divergence \cite{bengtsson2017geometry} between two probability distributions  $\bar p$ and $\bar q$ incrementally apart in time:
 \begin{equation}
     D_\mathrm{KL}(\bar p|| \bar q)= \sum_k p_k \log_2 \frac{p_k}{q_k}.
 \end{equation}
In the absence of continuous derivatives, we compute the KL divergence after each time step (kick), consistent with the physical intuition provided by Eq. \ref{fi}.  Upon increasing the particle number $N$, the energy space activity captured by the time-averaged $D_\mathrm{KL}(\bar p(n+1)||\bar p(n ))$ does not scale with $N$ as shown in Fig. \ref{fig:FI scaling}. What does this imply for charging despite the superlinear scaling of \(\Delta H_B^2\) and \(\Delta H_C^2\)?  This indicates that despite the rapid spreading of the state in the Hilbert space,  the probability flow between different energy levels is surprisingly slow. The scrambling in the Hilbert space is due to the chaotic nature of $H_C$ operator. However, rapid spreading in the Hilbert space is not the same as evolution in the energy eigenspace. Two orthogonal states in the Hilbert space can have identical energy distribution. The chaotic $H_C$ can scramble the state in the Hilbert space, but it need not lead to an effectively directed energy flow.

\begin{table}[b]
\caption{\label{tab:table1}Power bounds for parallel and global charging protocols based on Eq. \ref{fi_bound}.}
\begin{ruledtabular}
\begin{tabular}{lccc}
Charging protocol & $\Delta H_B^2$ & $I_E$ & $P$\\
\hline
parallel & $N$ & $N$ & $\leq N$\\
global & $N^2$ & $\sim N^0$ & $\leq N$\\
\end{tabular}
\end{ruledtabular}
\label{tab:table1}
\end{table}

With the improved bound in Eq. \ref{fi_bound}, as shown in Table \ref{tab:table1}, both the quantum and parallel charging protocols achieve the same power bound.  Note that the global protocol achieves the same $P \sim N$ through a large variance, $\Delta H_B^2 \sim N^2$ due to its entangled state. This compensates for a small Fisher information $I_E \sim \text{constant}$ from its single pulse.
However, using the bound in Eq. \ref{fi_bound}, the quantum advantage has not vanished but resides in the resources required to charge the battery to the same power as the classical case. For a global charger of the form $H_C^\#=\lambda \bigotimes_{i=1}^{N} h_C^i$, energy scales as $||H_C^\#|| \sim \lambda$ independent of $N$. The hardware complexity is also low, as a single global control device (e.g., lasers, microwave pulse, etc.)  is sufficient for charging. In parallel charging scenario, $H_C^{||} = \sum_{i=1}^N \lambda h_C^i$, the hardware complexity and the energy $||H_C^{||}||$ scale extensively $\sim N$, as independent chargers must be powered for each cell.

Does a significant energy space activity, given by $D_\mathrm{KL}(\bar p(n+1)||\bar p(n ))$, always mean better power? The answer is no, as illustrated by the following situations that arises because  $I_E$ is blind to the energy scale of the problem.
Consider a battery system-1 with two non-degenerate but very close energy levels with
battery Hamiltonian:
$H_B^1=\begin{pmatrix}
    0 &0 \\ 0&\epsilon
\end{pmatrix}$  with $\epsilon$ very small, and a battery-2 where $H_B^2=\begin{pmatrix}
    0 &0 \\ 0&1
\end{pmatrix}.$ The battery Hamiltonians differ only in their energy gap. We take the initial state in both systems to be $|1\rangle$ and charge them with the
Hamiltonian: $H_C = \lambda \sigma_x.$ The time evolution operator for both systems is given by 
\begin{equation}
U(t) = e^{-i H_C t} = \cos(\lambda t) I - i \sin(\lambda t) \sigma_x. \label{evolution}
\end{equation}
The state at time $t$ is:
\begin{equation}
    \ket{\psi(t)}= U(t) \ket{\psi_0}= \cos(\lambda t) \ket{1} - i \sin(\lambda t) \ket{0}. 
    \label{evolution}
\end{equation}
The population in each eigenstate is the same in both batteries and is given by $p_1(t)= \cos^2(\lambda t)$, and $p_0(t)= \sin^2(\lambda t).$ 
Then, expanding Eq. \ref{fi}, 
\begin{equation}
    I_E(t)= \frac{\dot p_0^2}{p_0} + \frac{\dot p_1^2}{p_1} =\lambda^2 \frac{\sin^2(2\lambda t)}{\sin^2(\lambda t)} + \lambda^2 \frac{\sin^2(2\lambda t)}{\cos^2(\lambda t)}. 
    \label{expanded FI}
\end{equation}
$I_E$ is the same regardless of the energy scale. The energy transfer is minimal, $P(t) \sim \epsilon$, in battery-1 because $\epsilon$ is small. However, $I_E$ can be large because the populations are changing rapidly. On the other hand, $P(t) \sim 1$ in battery-2, and $I_E$ at any given time is the same as that of battery-1.

Since $I_E$ only captures the change in the energy space distribution, it is blind to the direction of the energy flow.  Using the form of $I_E(t)$ in Eq. \ref{expanded FI}, we notice that only squares of the time derivatives of $p_k$ appear. Thus, $I_E$ is invariant with respect to the time reversal transformation,  $\dot{p}_k \rightarrow -\dot{p}_k$. Hence, $I_E$ fails to recognize charging from discharging.  

For example, consider a simple qubit battery $H_B = \frac{1}{2}\sigma_z$ undergoing Rabi oscillations driven by $H_C = \lambda \sigma_x$. As in the previous example, it starts in the fully discharged state $\ket{\psi_0}=\ket{1}$. From Eq. \ref{evolution}, one can see that the battery is fully charged (reaching $\ket{0}$) at $t=\pi/2\lambda$.  For $t > \pi/2\lambda$, the battery is in the discharging phase until $t=\pi/\lambda$, where $P(t)<0$. The functional expression for $I_E(t)$ is the same as the previous case, and is given by Eq. \ref{expanded FI}.
From Eq. \ref{expanded FI}, it can be verified that
\begin{equation}
    I_E(t)= I_E\left(\frac{\pi}{\lambda} -t\right),
\end{equation}
where $0\leq t\leq \pi/2\lambda$. Thus, at time $t$ (charging) and time $(\pi/\lambda - t)$ (discharging), $I_E$ is identical, but the power $P(t)$ has the same magnitude and opposite sign. This example indicates that the upper bound containing $I_E$ can be misleading if used in isolation.  A high $I_E$ can indicate efficient charging or rapid, wasteful self-discharge. One cannot distinguish a healthy battery from a faulty one using $I_E$ alone.

Furthermore, $I_E$ can be large due to dynamics within a degenerate subspace. Consider a battery Hamiltonian with a degenerate subspace: \begin{equation}
    H_B= \begin{pmatrix}
    0&0&0\\
    0&0&0\\
    0&0&1
\end{pmatrix}
\end{equation}
$H_B$ has two degenerate ground state levels, $\ket{g_1}$ and $\ket{g_2}$, with energy $E=0$, and an excited level $\ket{e}$ at $E=1$. Now, consider a charging Hamiltonian that only couples the two degenerate levels: $H_C= \lambda(\ket{g_1}\bra{g_2} +\ket{g_2}\bra{g_1}).$  For an initial state $ \ket{\psi_0}=\ket{g_1}$, whose dynamics according to $H_C$ is completely confined to the degenerate subspace, the Hamiltonian can be represented as a $2 \times 2$ matrix in the  $\{\ket{g_1} , \ket{g_2}\}$ basis:
\begin{equation}
H_C= \lambda \begin{pmatrix}
    0&1 \\1&0
\end{pmatrix}.
\end{equation}
Then the evolved state is $\ket{\psi(t)}= \cos(\lambda t) \ket{g_1} -i\sin(\lambda t)\ket{g_2}$.
Again, in this case, $I_E(t)$ is the same as Eq. \ref{expanded FI}:
\begin{equation}
    I_E(t)= \frac{\dot p_{g_1}^2}{p_{g_1}} + \frac{\dot p_{g_2}^2}{p_{g_2}} = \lambda^2 \frac{\sin^2(2\lambda t)}{\sin^2(\lambda t)} + \lambda^2 \frac{\sin^2(2\lambda t)}{\cos^2(\lambda t)}. \label{toggle}
\end{equation}
Then, in general $I_E > 0$, even though the battery state is only hopping between two degenerate levels. There is no charging taking place in this case, and both the $\langle H_B\rangle_{\psi(t)}$  and $P(t)=0\: ~\forall ~t$.  However, $I_E(t)$ fails to recognize this, which is a critical flaw. The activity measured by $I_E$ is entirely wasted on population changes within a degenerate subspace, resulting in no net energy transfer to the battery.

Another issue with Fisher Information is that it can diverge  when the power is zero, typically at points where the population of an energy level approaches zero. In the Rabi oscillation example discussed above, consider $t= \pi/2\lambda,$ where $p_0=1$ and $p_1=0.$ The battery is fully charged and the energy is maximum, with $P(t)=\frac{dE}{dt}=0.$ At this instant, there is no charging taking place. However, $I_E(t)$ diverges at this point as $p_0 \to 0$ in Eq. \ref{expanded FI}. Thus, $I_E$ is unstable at the inflection points, where the system changes from charging to discharging or vice versa.

Among the various spin-battery systems analyzed in \cite{julia2020bounds}, the only occasion where the power bound involving $I_E$ shows superlinear scaling ($\sim N^{1.5}$) is in the Dicke model of quantum battery in the regime of strong coupling between the charger and the battery. The Hamiltonian is given by
\begin{equation}
    H_{\text{DK}} = J_z + \hat{a}^\dagger \hat{a} + \frac{2\lambda}{\sqrt{N}} J_x (\hat{a}^\dagger + \hat{a}).
\end{equation}
Here, $J_x$ and $J_z$ are collective spin operators, and  $\lambda$ is the coupling strength between the spins and the cavity.  In the strong coupling regime, $\lambda=0.5$.  The operators $\hat{a}^\dagger$ and $\hat{a}$ are the creation and annihilation operators of cavity photons. In this case, the charging proceeds by gradually occupying many intermediate states \cite{julia2020bounds}. This gradual energy transfer via intermediate states is effected by $J_x(\hat{a}+\hat{a}^\dagger)$ term. Yet, this system has a superlinear power bound ($\sim N^{1.5}$). This is due to higher energy space activity involving multiple energy levels. The authors correctly emphasize that this bound is far from saturated in this case, and the actual power achieved remains linear in $N$.

Now, let us compare with the charging performance of the Dicke battery with a {\it global} charger. For $H_B= \frac{1}{2}\sum_{i=1}^{N} \sigma_z^i$ and the fully discharged initial state $\ket{\psi_{initial}}=\ket{0}^{\otimes N},$ a global charger of the form $H_C= \lambda \bigotimes^N \sigma_x^i$
achieves the fully charged state $\ket{\psi_{final}}=\ket{1}^{\otimes N}$ without transiting through the intermediate energy states \cite{julia2020bounds}. The global charging is efficient as it couples the lowest energy state only with the highest energy state, and the system jumps directly between the energy extremes. In the global charging case, $I_E$ is independent of the system size, and the power bound scales linearly with $N$.

Thus, comparing these two scenarios for the Dicke model, the upper bound involving $I_E$  fails to reflect the rate of energy transfer in Dicke battery. Global charging has the most efficient energy transfer, and a physically meaningful upper bound should reflect that. However, the bound incorrectly signals a possible quantum advantage in power scaling for the Dicke battery, which is intangible. 

Despite these issues that can potentially inflate the power bound, we observed that replacing $\Delta H_C^2$ with $I_E$ already eliminates an order of magnitude advantage in charging power in the kicked spin battery. A similar loss of superlinear advantage is observed in various other spin models of quantum batteries \cite{julia2020bounds}.

A second potential issue is with the role of entanglement, represented by $\Delta H_B^2$, in achieving tangible quantum advantage. Traditionally, $\Delta H_B^2$ is central for characterizing quantum advantage, given that the other term in the product (either $\Delta H_C^2 $ or $I_E$) is properly taken into account. For instance, extensive $\Delta H_C^2$ along with a quadratic $\Delta H_B^2$ is considered a smoking gun of genuine quantum advantage\cite{rossini2020quantum}. This is because the superlinear scaling $P \sim N^{1.5}$ stems from an enhanced $\Delta H_B^2$, which is quantum in origin. The quadratic $\Delta H_B^2$  is interpreted to be linked to an uncharged battery state traversing shortcuts in the Hilbert space via highly entangled states to the charged state. However, this interpretation of $\Delta H_B^2$ is problematic.  Entanglement does not always translate into useful charging. For instance, extensive $\Delta H_C^2$ along with a superlinear $\Delta H_B^2$ in kicked spin model and in the Dicke quantum battery in the strong coupling regime\cite{julia2020bounds} do not yield tangible quantum advantage. Hence, mere presence of large entanglement is not a sufficient condition for quantum advantage in charging \cite{gyhm2024beneficial}.

Furthermore, a large $\Delta H_B^2$ could arise from the mixedness of the battery state, rather than from internal correlations between the battery cells. Such a situation occurs if the entanglement between the battery and charger persists in the final state,  diminishing the quality of charging, as the stored energy in the reduced battery state is now probabilistic, with significant variance \cite{julia2020bounds}. 


The preceding discussions can be summarised as follows: as we show using the spin-chain battery model in Eqs. \ref{eq:battery1}-\ref{charger}, even the well-meaning power bounds may not be tight in a range of physical situations. Contribution to superlinear scaling of the bound can arise from both $I_E$ as well as $\Delta H_B^2$ (battery) terms. While a superlinear power bound signals a possible quantum advantage in principle, it may not be tangible or even exist in practice. This platform-independent analysis has implications for experimental quantum battery realizations. It can be misleading if quantum advantage is claimed only based on power bounds. Further analysis must be performed to verify if $I_E$ reflects useful charging process.


In addition to the theoretical problems discussed, $I_E$ is also difficult to track in experiments. It requires the knowledge of individual populations $p_k$ for all relevant energy levels labelled by $k$. The number of eigenvector projections scale exponentially with system size. Furthermore, computing $I_E$ at any given instant involves a time derivative, and state characterization at nearby points is also necessary. 

 {We note that there are other power bounds in the literature -- based on the fluctuations of the free energy operator \cite{garcia2020fluctuations}, and one based on the quantum speed limit \cite{campaioli2017enhancing}. The key message of this paper \begin{equation}
    \textrm{Higher upper bound } \neq \textrm{More quantum advantage}, \nonumber
\end{equation}   holds as a note of caution for quantum advantage claims based  on any power bounds in general.} An accurate, { experiment-friendly} power bound that reflects the actual energy flow, without the problems discussed in the previous sections, is an interesting future direction. A complete theory needs to incorporate how the geometry of the battery's energy spectrum ($H_B$) and the nature of the dynamics ($H_C$) interact to determine the useful, non-wasteful part of $I_E$ that {\it actually contributes} to charging. {It should also be able to isolate useful entanglement from the total entanglement present. } 
 {Moreover, entanglement is a necessary but not  sufficient condition for quantum enhanced charging. A recent work shows that power is not an entanglement monotone \cite{gyhm2024beneficial}, but depends on the state’s geometric structure in Hilbert space to align with an optimal driving pathway. This  calls for a careful study of the actual resource for faster charging, along with a systematic classification of states for their practical use. } 
 \section*{acknowledgements}
 S.PG acknowledges the I-HUB quantum technology foundation (I-HUB QTF) at IISER Pune for financial support (I-HUB/PDF/2024-25/010/52).
 \section*{DATA AVAILABILITY}
The data that support the findings of this study are available from the corresponding author upon reasonable request.
\FloatBarrier
\bibliography{ref}
\bibliographystyle{unsrt}
\end{document}